\documentclass[twocolumn,]{aastex631}
\makeatletter
\let\frontmatter@title@above=\relax
\makeatother
\usepackage[utf8]{inputenc}
\usepackage[T1]{fontenc}

\begin{document}

\title{SPICE: Scintillation Pipeline for Interferometric Candidate Extraction}

\author{Jitendra Salal}
\affiliation{National Centre for Radio Astrophysics, Post Bag 3, Ganeshkhind, Pune, 411007, India}

\author{Shriharsh Tendulkar}
\affiliation{Tata Institute of Fundamental Research Mumbai,India}
\affiliation{National Centre for Radio Astrophysics, Post Bag 3, Ganeshkhind, Pune, 411007, India}

\author{Visweshwar Ram Marthi}
\affiliation{National Centre for Radio Astrophysics, Post Bag 3, Ganeshkhind, Pune, 411007, India}

\begin{abstract}

We present Scintillation Pipeline for Interferometric Candidate Extraction (SPICE) an automated CASA‑based pipeline developed to identify pulsar candidates in Giant Metrewave Radio Telescope (GMRT) and upgraded GMRT (uGMRT) data through their diffractive interstellar scintillation signatures. SPICE integrates flagging, calibration, imaging, and classification, with robust RFI excision, iterative self‑calibration with dynamic reference antenna selection, source detection using \texttt{PyBDSF}, and classification based on our earlier development of scintillation-based visibility correlation searches. SPICE is available publicly on github https://github.com/G2-SL/SPICE and is archived on Zenodo \citep{spice_salal2026}. We applied SPICE to archival datasets from both legacy GMRT and uGMRT. The pipeline successfully recovered known pulsars such as PSR\,J0437$-$4715, PSR\,B0450$-$18, and PSR\,B0329+54, yielding scintillation parameters consistent with expectations. Non‑detections in some scans highlight the influence of pervasive RFI, the dependence on the reference antenna, and the intrinsic variability of the scintillation properties. SPICE complements time‑domain searches by enabling reproducible scintillation‑based candidate identification in interferometric data. Its application to the GMRT archive opens a pathway for discovering compact variable sources and expanding pulsar searches beyond time-domain searches.

\end{abstract}

\section{Introduction} \label{sec:intro}

The GMRT, operational since 2002, covered five frequency bands between 120 and 1450\,MHz. The recent upgrade replaced narrow-band receivers and backends with broadband systems, increasing the maximum bandwidth to 400\,MHz. The uGMRT now operates in four frequency bands: 120$-$250\,MHz (Band\,2), 250$-$500\,MHz (Band\,3), 550$-$850\,MHz (Band\,4), and 1000$-$1460\,MHz (Band\,5). The GMRT data archive contains observations from two primary correlators: the GMRT Software Backend (GSB) and the GMRT Wideband Backend (GWB). GSB data offer bandwidths of 16.66 or 33.33\,MHz, while GWB data, available since 2021, provide up to 200\,MHz. Spectral resolution ranges from 128 to 16,384 channels, depending on observational requirements. The corresponding channel width depends on the selected bandwidth. For example, with 16\,MHz bandwidth, a channel width of 125\,kHz corresponds to 128 channels, while with 200\,MHz bandwidth, a channel 
width of 192\,kHz corresponds to 1024 channels. At the finest resolution, 
16,384 channels over 200\,MHz bandwidth with a channel width of $\sim$12\,kHz. Individual scans typically last 5\,min to 1\,h, with subintegration times of 8\,s (adjustable between 2\,s and 32\,s). Most archival observations target regions near the Galactic plane. 

Radio interferometric data from the GMRT require standard post-processing steps such as flagging, calibration, and inverse Fourier transformation to produce sky images. Traditionally, these data were analyzed interactively using the NRAO Astronomical Image Processing System (AIPS). This approach can become cumbersome for large datasets or when near real-time analysis is needed. To address these challenges, automated pipelines have been developed for GMRT observations. Source Peeling and Atmospheric Modeling (SPAM; \citealt{intema2009,spam}) is an AIPS$-$based system originally designed for the legacy GMRT, primarily for lower$-$frequency observations ($<$1\,GHz). However, SPAM does not produce calibrated visibility data, which are essential for constructing dynamic spectra. More recently, CAPTURE \citep{kale2021}, a CASA-based pipeline customized for uGMRT, has been introduced. CAPTURE performs flagging, calibration, and imaging of the target field, but it is optimized for producing a single image from all scans of a target. 

Our work builds on these developments by constructing a dedicated scintillation pipeline to detect compact radio sources that exhibit variability due to diffractive interstellar scintillation (DISS). In interferometric images, pulsars typically appear as compact sources with steep spectral indices \citep{backer}, some of them with highly circularly polarized emission \citep{wang2022}, and time variability caused by scintillation. CAPTURE is well-suited to generate high-quality images from uGMRT data, but our scintillation analysis requires scan-by-scan calibration and imaging, which is crucial because both the noise characteristics and scintillation parameters vary over time. This approach enables us to identify pulsar candidates through their scintillation signatures in the interferometric data and subsequently confirm them via time-domain searches. Our scintillation-based visibility correlation searches (SVCS) technique cross-correlates dynamic spectra derived from visibility data to measure scintillation bandwidths and timescales, as described in our earlier work \citep{salal} (hereafter S24). In a follow-up study, we used \texttt{PsrPopPy} \citep{psrpoppy} to simulate pulsar populations and established the operational boundaries of SVCS in terms of DM ranges and duty cycle constraints, while quantifying its detection efficiency relative to time-domain searches \citep[][hereafter S26]{salal2026}. 

In this paper, we present a scintillation pipeline for GMRT archival data to identify known pulsars using SVCS. Section~\ref{sec:spice} describes the structure of the scintillation pipeline SPICE, Section~\ref{sec:result} presents examples of GMRT datasets processed with SPICE, Section~\ref{sec:discussionspi} discusses its limitations, and Section~\ref{sec:conclusionspi} provides a summary.

\section{SPICE} \label{sec:spice}

\begin{figure*}
\includegraphics[width=\textwidth]{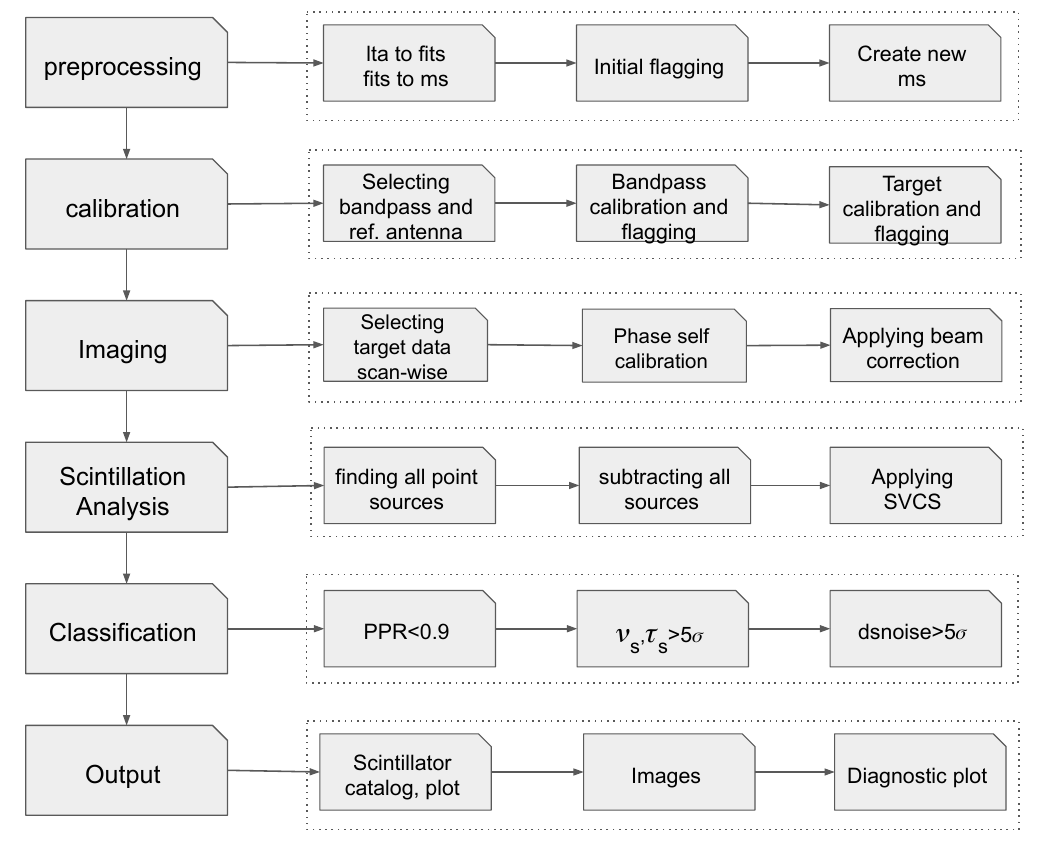}
\caption{The structure of the scintillation pipeline SPICE. The left column shows the broad workflow, beginning with the LTA or FITS file. The row in front of each block in the left column indicates the specific process occurring at that stage. At the end of the pipeline run, the outputs include the radio image, diagnostic plots, and the scintillator catalog with corresponding plots. Here, dsnoise is the dynamic spectrum noise, PPR is the peak-to-peak ratio, $\nu_s$ and $\tau_s$ are the scintillation bandwidth and timescale.
}
\label{fig:spice}
\end{figure*}

SPICE is a scintillation pipeline built on CASA~6.3, the Python-based version of CASA. GMRT data are provided in LTA or FITS format, both of which contain interferometric visibilities (UVFITS) rather than pulsar time‑series (PSRFITS). If the data are in LTA format, we first translate them into FITS and then convert the FITS files into measurement sets (MS) using the CASA task \texttt{importgmrt}. The structure of SPICE is shown in Figure~\ref{fig:spice}.

\subsection{Flagging} \label{sec:style}

Ideally, no flagging would be necessary, but in practice, it is essential due to radio frequency interference (RFI). GMRT data can be corrupted by a variety of factors, including instrumental failures, bad antennas, ionospheric scintillation, and RFI from both natural and man-made sources. Such interference may be narrow-band and continuous, originating from nearby electronics such as Wi-Fi routers or control systems, or broadband and intermittent, caused by sources such as lightning strikes, mobile communication towers, or aircraft transponders. Additional sources include electric sparks, emissions from satellite communication bands, and faulty baselines. To mitigate its effects, data are flagged prior to further processing. In CASA, this is typically done using the task \texttt{flagdata}, which identifies and removes corrupted visibilities from the dataset. 

We began by flagging visibilities with amplitudes greater than 100\,Jy or equal to 0\,Jy. We then removed data from permanently contaminated frequency ranges for GMRT, when present, namely 360–379.6\,MHz, 486–493.6\,MHz, 764.6–769.1\,MHz, and 880.8–885.6\,MHz. In addition, we flagged the first channel of each scan, as well as the first and last 2\,s of data. Target scans containing fewer than 25 time samples were flagged, as such datasets do not provide sufficient temporal coverage for reliable cross-correlation. Similarly, calibrator scans with fewer than two time samples were excluded. 

To flag time samples from each scan, we calculate the mean and standard deviation of the visibility amplitude at each time sample. We then compare these values against the distribution across all time samples in the scan and flag those whose mean or standard deviation exceeds \(3\times\) the median absolute deviation (MAD) of the mean or standard deviation. 

\[
\mu_t > 3 \times \mathrm{MAD}(\mu_t) 
\quad \text{or} \quad 
\sigma_t > 3 \times \mathrm{MAD}(\sigma_t),
\]

\begin{equation}
\mathrm{MAD}(X) = \frac{1}{N}\sum_{i=1}^{N} \left| X_i - \mathrm{median}(X) \right|.
\end{equation}

If a scan is left with fewer than two time samples after this step, the entire scan is flagged. To flag channels from each scan, we similarly compute the mean and standard deviation of the visibility amplitude for each channel and flag those whose mean or standard deviation exceeds \(3\times\) MAD of the mean and standard deviation. 

To identify bad antennas, we selected the central 40\% of channels and calculated the mean and standard deviation of the visibility amplitude for each antenna in each calibrator scan, then find the antennas that have a mean or standard deviation greater than 2$\times$ MAD of the mean or standard deviation in each scan. Antennas that exceed 2$\times$ MAD in every calibrator scan are considered bad antennas. 

The central baselines (i.e., baselines formed between antennas in the central square) are strongly affected by the RFI and are therefore excluded from the analysis. Since our focus is on compact point sources, flagging these short baselines has little impact on the results as they are primarily important for imaging extended structures. In practice, excluding them leads to only a slight loss of sensitivity while significantly improving image quality. For example, in GMRT observation 1665 at 333\,MHz on 2004 December 12, inclusion of the central baselines increased the image noise to the point that PSR\,B0450$-$18 became undetectable. Consequently, all 91 central baselines, together with those involving bad antennas, are flagged. 

We require only Stokes$-$I images, so we consider only the RR and LL correlations. Using the \texttt{split} task in CASA, we extract the relevant data from the original measurement set, excluding all bad baselines, extreme flagged time samples, extreme flagged channels, and the RL and LR cross-correlations. The resulting measurement set is smaller in size, takes less time to analyze, and requires less memory. We then ran the \texttt{tfcrop} task on the visibility file. 

\subsection{Calibration}
Calibration in radio imaging is essential because the raw signals from the telescopes are distorted by instrumental imperfections, atmospheric effects, and electronic noise, making uncorrected images unreliable. It ensures that the data reflect the true sky by correcting for antenna sensitivity variations, frequency-dependent gain, and time-varying delays caused by the ionosphere or electronics. Calibration also converts raw voltages into physical brightness units, enabling accurate flux measurements, comparisons between different instruments, and long-term monitoring of sources. By removing spurious patterns and sharpening resolution, it improves image fidelity and allows the detection of faint objects. 

\textbf{Bandpass calibrator:} 
We first identify a bandpass calibrator scan from among all the available calibrator scans. SPICE selects the calibrator with the largest number of visibility data points. If that scan is flagged after the initial round of calibration and flagging, SPICE discards it and instead selects the calibrator with the next largest number of data points, repeating the calibration and flagging process. This procedure continues until a bandpass calibrator with the largest unflagged visibility dataset is found. The remaining calibrator scans are then used as phase calibrators.

\textbf{Reference antenna:} 
For calibration, a reference antenna is also required. To select it, we calculated the circular standard deviation of the visibility phases of the GMRT arm antennas for the bandpass calibrator and chose the antenna with the smallest circular standard deviation. The circular standard deviation for variable $\theta$ is defined as
\begin{equation}
    \sigma=\sqrt{-2ln|<e^{i\theta}>|}
\end{equation}

\textbf{Bandpass and phase calibration:} 
We did not perform any flux scaling because most pulsar interferometric datasets lack a flux calibrator, and our analysis focuses on scintillation signatures, which do not require the absolute flux density of the source. Therefore, the flux density we obtain is relative. We performed two cycles of phase, delay, bandpass, and gain calibration on the bandpass and phase calibrators. Each cycle was followed by flagging using the \texttt{flagdata} mode \texttt{tfcrop} and \texttt{rflag}. Subsequently, we removed all baselines that were flagged in every calibrator from the visibility file, which helped reduce computation time in later stages of analysis. 

After calibrating the bandpass and phase calibrators, SPICE performed three cycles of bandpass and phase calibration on the target scans, each followed by flagging with \texttt{tfcrop}, \texttt{rflag}, and 5$\sigma$ phase flagging of the calibrators. We also applied \texttt{flagdata} in the \texttt{extend} mode to flag all time samples and channels that had already been flagged more than 80\%. At the end of the calibration, SPICE saves the calibration tables and diagnostic plots for each calibrator. These plots show the calibrated visibility amplitude as a function of time, frequency, UVdist, and calibrated visibility phase. 

\subsection{Imaging}
The imaging process applies Fourier transforms to calibrated visibility data to produce radio images, enabling the detection and analysis of astrophysical sources. SPICE generates these radio images using the \texttt{tclean} task in CASA. The \texttt{cell} parameter is set to be frequency dependent, so that the pixel size automatically matches the angular resolution at the central observed frequency, calculated using the Rayleigh criterion,  
\begin{equation}
    \theta= \frac{\lambda}{D} \ \mathrm{rad},
\end{equation}  
where for GMRT, $D$ (maximum baseline length) is 25\,km and $\lambda$ is the wavelength corresponding to the central frequency of observation. The \texttt{imsize} is set to [3200, 3200], corresponding to a field of view of approximately 20\,arcmin, and the pixel size is $\theta/5\sim$ 1.77\,arcsec at 1.4\,GHz, giving an image dimension of 3388$\times$3388 pixels. To take advantage of the optimized FFT routines used internally by CASA, it is recommended that the number of pixels along each axis be even and factorizable by small primes (2, 3, and 5). We used 3200 pixels (close to 3388), independent of frequency. The \texttt{threshold} parameter is adaptive, updates dynamically based on the RMS noise of intermediate images, and terminates the cleaning process when residual fluctuations approach the noise floor. 

SPICE uses the CASA task \texttt{tclean} with 
\texttt{deconvolver='hogbom'}, which applies the classic CLEAN algorithm 
to iteratively subtract point-source components from the dirty image, 
producing a deconvolved sky model. For visibility weighting, we adopt 
\texttt{weighting='briggs'} with \texttt{robust=0.5}, which balances 
sensitivity and resolution by adjusting how visibilities are weighted 
between natural (sensitivity-optimized) and uniform (resolution-optimized) 
schemes. We set \texttt{niter=5000} to allow sufficient CLEAN iterations 
for convergence. After the initial radio image, SPICE applies two rounds 
of phase-only self-calibration to refine antenna gain solutions. The 
solution interval is equal to the subintegration time, correcting 
short-timescale phase variations. The data are then re-imaged using the 
updated calibration tables, reducing phase errors and improving source 
coherence.

SPICE corrects the primary beam response using the CASA task \texttt{widebandpbcor}, which accounts for the frequency-dependent beam shape—broadening at lower frequencies, and adjusts flux densities to compensate for sensitivity losses at the field edges. The resulting images have uniform noise properties and corrected primary-beam attenuation on a relative flux scale.

\subsection{Classification}

Compact radio sources exhibiting DISS often correspond to pulsars or other Galactic objects with angular sizes smaller than the angular scale of interstellar turbulence in the ionized interstellar medium. These sources appear as unresolved objects in radio images, with flux densities that vary temporally and spectrally due to scintillation. 

To identify such sources, we use the Python Blob Detection and Source Finder (\texttt{PyBDSF}) package \citep{pybdsf}, which decomposes the radio image into discrete emission regions (“blobs”) and fits Gaussian components to each significant peak. We configure \texttt{PyBDSF} with \texttt{$thresh_{isl}$} = 3.0 and \texttt{$thresh_{pix}$} = 5.0, corresponding to detection thresholds of 3$\sigma$ for island boundaries and 5$\sigma$ for individual pixels, thus minimizing false positives from noise fluctuations. The parameter \texttt{$atrous_{do}$} is set to \texttt{False} to disable multiscale decomposition and favor point-source detection over extended emission. We set \texttt{$rms_{box}$} = (240, 70) to define a sliding box to measure local noise variations throughout the image, avoiding overestimation in regions containing bright sources. The resulting catalog includes positions, peak fluxes, integrated fluxes, and major and minor axis dimensions for all detected components. We then filter candidates by selecting sources with a peak-to-integrated flux ratio greater than 0.8, indicative of unresolved emission characteristic of scintillating point sources, and with a major axis smaller than 1.5 times the synthesized beam FWHM to exclude extended or resolved structures. 

After finding the compact radio source in the image, we select a source from the catalog and phase shift it to the phase center, then form the visibility dynamic spectra using the visibility addition method \citep{salal}. We then cross-correlate the dynamic spectra and fit the 2D-Gaussian model to calculate the scintillation parameters. We also fit 1D Gaussians to the cross-correlation at zero frequency lag along the time-lag axis and at zero time lag along the frequency-lag axis to get the scintillation timescale and bandwidth with error. We calculated the noise in the dynamic spectra, modulation index, and correlation SNR. 

We also calculate the peak-to-peak ratio (PPR), which is defined as the ratio of the peak of the off-center cross-correlation to the peak near the center with a 7 pixel radius. A truly scintillating radio source has PPR close to zero because it gives a very high peak in the central region and a very low peak outside it. To classify the source as a scintillating source, we impose constraints on PPR, which should be less than 0.9, the scintillation timescale and bandwidth, which should be greater than 5$\sigma$, and dynamic spectrum noise, which should be greater than 1$\sigma$ among all sources in the scan.

\section{Results} \label{sec:result}
\begin{figure*}
\includegraphics[width=\textwidth]{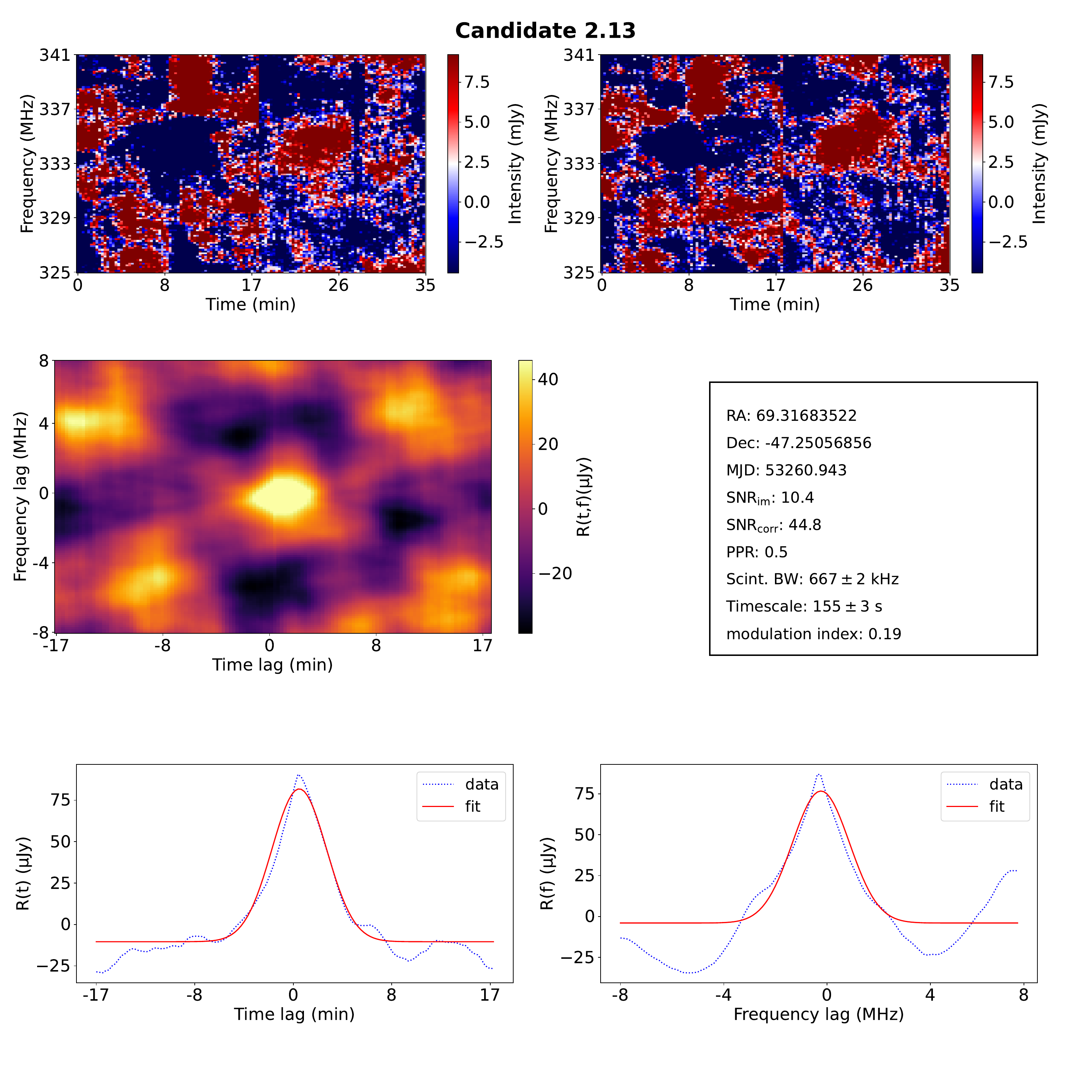}
\caption{Scintillator plot showing the dynamic spectra, cross-correlation of dynamic spectra, one-dimensional cross-correlation along time and (frequency) axis at zero frequency (time) lag, along with one-dimensional Gaussian fit. In Candidate 2.13, the first and second number corresponds to the scan and compact sources order given by \texttt{pyBDSF}, respectively.}
\label{fig:scint_plot}
\end{figure*}

SPICE is fully automated and requires an LTA or FITS data file from a GMRT observation. To run SPICE, we use the command line \texttt{python spice.py /path/to/folder}, where the specified folder must contain the LTA or FITS file. The pipeline outputs diagnostic plots, calibration tables, and radio images of the target scans, along with a log file listing all classified scintillators with their RA, Dec, SNR, relative flux density, scintillation bandwidth, and timescale (with errors), modulation index, MJD, channel width, central frequency, bandwidth, integration time, subintegration, correlation SNR, and project code. In addition, it generates a plot showing the dynamic spectrum, its cross-correlation, and one-dimensional projections along time (frequency) at zero frequency (time) lag, each fitted with a one-dimensional Gaussian as shown in Figure~\ref{fig:scint_plot}. We applied SPICE to two GMRT datasets: one from legacy GMRT (12\,September\,2004, project code 06YGA01) and the other from uGMRT (31\,July\,2017, project code 32\_089). 

\begin{figure*}
\centering
\includegraphics[width=0.5\textwidth]{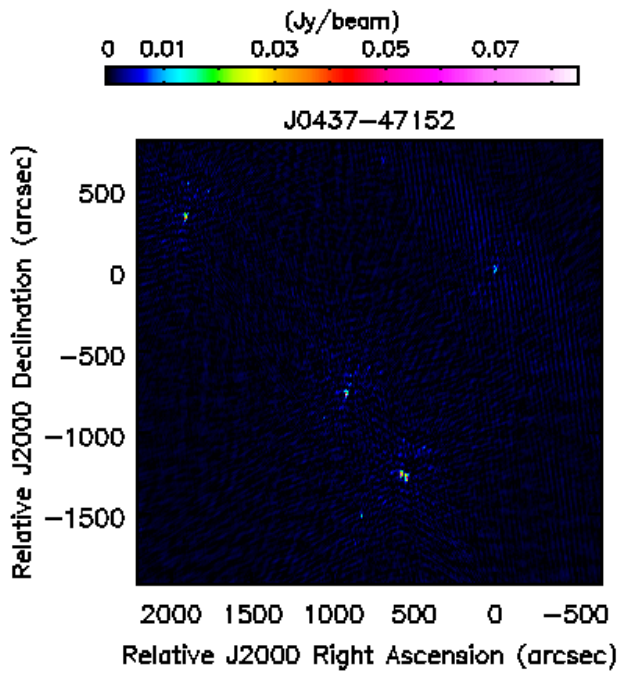}
\caption{Radio images of PSR\,J0437$-$4715 produced with SPICE. The pulsar has an image SNR of 13.6, observed at 333\,MHz with a 16\,MHz bandwidth and an on‑source integration time of 26\,min.
}
\label{fig:radio_image_j0437}
\end{figure*}

The legacy GMRT dataset (observation number 1665) has an integration time of 16.78\,s, observed at the central frequency $f = 333$\,MHz with 128 channels. It includes seven scans of known pulsars: PSR\,J0437$-$4715 (DM = 2.6) was observed in two scans of 26.25 and 35.79\,min, both yielding pulsar detections via SVCS for the scan with duration 26\,min, the scintillator plot is shown in Figure~\ref{fig:scint_plot} and the radio image is shown in Figure~\ref{fig:radio_image_j0437}. PSR\,B0450$-$18 (DM = 39.9) was observed in two scans of 10.52 and 29.08\,min, both resulting in detections. PSR\,B0740$-$28 (DM = 73.7) was observed in a single 133.34\,min scan and detected. NGC\,1851, a globular cluster known to host multiple pulsars, was observed in two scans of 64.31 and 63.19\,min, without detections. The observation data (1665) are stored in a 2.2\,GB FITS file. On a workstation with 36 cores and 755 GB of memory, calibration required approximately 27\,min, while imaging the first scan of PSR\,J0437$-$4715 took approximately 7\,min. The imaging time scaled with scan duration, with NGC\,1851 (scan duration $>1$\,h) requiring around 17\,min. Extracting the scintillation parameters from each dynamic spectrum for PSR\,J0437$-$4715 took approximately 7.5\,s per source. The complete SPICE run on this dataset was completed in roughly 1.8\,h. 

\begin{figure*}
\centering
\includegraphics[width=0.5\textwidth]{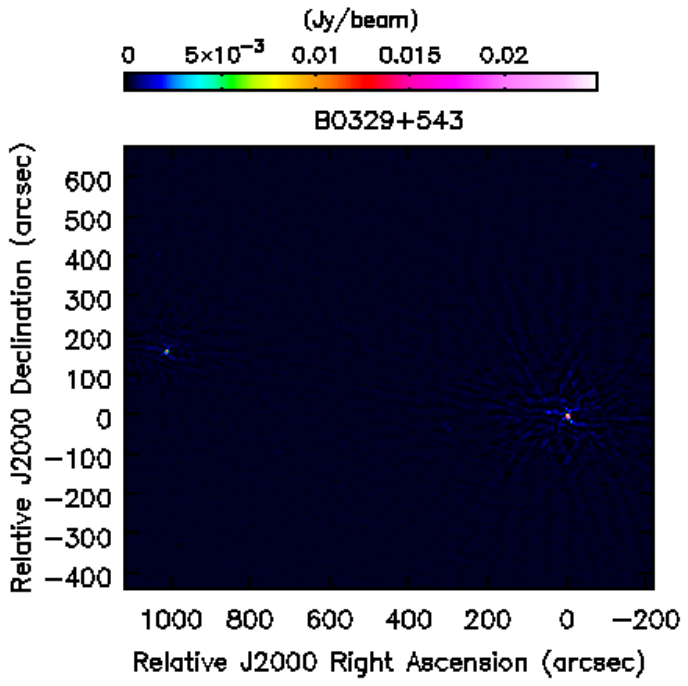}
\caption{Radio image of PSR\,B0329+54 produced with SPICE. The pulsar has an image SNR of 35.3, observed at 400\,MHz with a 200\,MHz bandwidth and an on‑source integration time of 25\,min. 
}
\label{fig:radio_image_B0329}
\end{figure*}

The uGMRT dataset (observation number 9622) has an integration time of 5.37\,s, observed at the central frequency $f = 400$\,MHz with 2048 channels. It includes five scans of known pulsars with an average duration of 25\,min each. The radio image is shown in Figure~\ref{fig:radio_image_B0329}. PSR\,B0329+54 (DM = 26.76) yielded pulsar detection in SVCS. PSR\,J0218+4232 (DM = 61.3) was detected in the radio image but not through SVCS. The NE2001 model predicts a scintillation bandwidth of 40\,kHz and a timescale of 170\,s, which should allow detection, albeit with low modulation because the channel width 200\,kHz is larger than the 40\,kHz. The non-detection may be due to inaccurate predictions of scintillation bandwidth and timescale; moreover, scintillation parameters vary with time due to refractive interstellar scintillation, which can result in a very low effective scintillation bandwidth. PSR\,J0437$-$4715 (DM = 2.6) yielded detections in SVCS. PSR\,J0613$-$0200 (DM = 38.78) was detected in the radio image but showed a PPR of 0.8; most other sources in this scan exhibited similar values, suggesting pervasive RFI in this scan that caused many sources to appear scintillating. PSR\,J0621+1002 (DM = 36.55) showed no detection in the radio image. The noise near the pulsar location in the radio image is very high. The observation data (9622) are stored in a 27\,GB FITS file. Calibration required approximately 5.1\,h, while imaging the first scan of PSR\,B0329+54 took approximately 2.6\,h. Extracting the scintillation parameters from each dynamic spectrum in the field PSR\,B0329+54 took approximately 28\,s per source. The complete SPICE run on this dataset was completed in roughly 23\,h. 

We applied SPICE to more than 100 GMRT archival datasets, each containing 
on average $\sim$500 compact sources. In total, we analyzed about 50,000  compact sources and detected approximately 350 scintillating sources, of which 240 are known pulsars. The remaining 110 scintillating sources are classified as pulsar candidates. A detailed study of these candidates and their application to probing the structure of the interstellar medium will be presented in an upcoming paper. We also identified one scintillating source that exhibits features similar to those of a known pulsar but is not associated with any catalogued pulsar. Follow-up observations suggest that this source could be a pulsar. We are continuing to monitor it and will publish our findings in a subsequent paper.

\section{Discussion}\label{sec:discussionspi}

\subsection{Limitation of SVCS}
We cannot measure the true scintillation timescale and bandwidth because of the limitations in frequency and time resolution of the telescope. The measured scintillation timescale and bandwidth are approximations to the true values. In the case of GMRT, we can measure the true scintillation timescale due to a time resolution of 2\,s as discussed in S26, and the simulation shows that the minimum scintillation timescale is 2\,s. The simulation also shows that the minimum scintillation bandwidth is in Hz, whereas the GMRT frequency resolution is in kHz, so it is difficult to determine the true scintillation bandwidth for all scintillating sources. If we use baseband data for high-frequency resolution to measure the true scintillation bandwidth, the sensitivity decreases because the sensitivity of the detection is directly proportional to the square root of the measured scintillation bandwidth. 

To calibrate the radio visibility data, we are required to set a reference antenna. The resulting radio image and the further measurements of the scintillation parameters are highly dependent on the reference antenna. If the reference antenna suffers from unstable gains or RFI contamination, 
calibration errors can propagate across the array, sometimes causing bright sources to disappear from the image. For example, in a GMRT archive project (project code: 1539), scan 18 (field name J0437-4715) has a PSR\,J0437$-$4715 is not detected in the radio image, when the reference antenna is set to 'E01','E03','W04', or 'W01'. In practice, reference antennas are  typically chosen from the central square antennas with stable electronics and good sensitivity, and consistency across scans is recommended. Similar challenges in 
reference antenna selection have been noted in other GMRT calibration studies \citep{prasad2012, intema2017}, where careful choice was required to ensure robust phase solutions.

\subsection{Interference and instrumental effects}
In the SPICE calibration stage, flagging is performed to mitigate RFI but does not eliminate all instances. We have demonstrated various methods to reduce the effects of RFI and instrumental variations, such as flagging the central square baselines, bad antennas, bad baselines, bad channels, etc.; however, the possibility of baseline-dependent variations manifesting as scintillation cannot be ruled out. In such cases, the scintillation parameters of many sources in the image may appear similar and interdependent. We have also shown that the impact of affected baselines can be reduced by flagging short baselines that are particularly prone to contamination. 

As discussed in S24, the baseline-dependent RFI can modify the scintillation parameters in the autocorrelation, forcing us to rely on cross-correlation at the cost of reduced $\mathrm{SNR_{cor}}$. Additional forms of RFI are also present in the dynamic spectrum, such as RFI caused by ripples in frequency and time, which manifest themselves as repeating patterns in the correlation. Dynamic spectra derived from different baseline subsets yield scintillation parameters that are broadly consistent but statistically distinct. However, baseline splitting cannot fully remove baseline dependence since the resulting dynamic spectra still share common antennas.

\section{Conclusion}\label{sec:conclusionspi}

The construction of a radio image is essential for detecting pulsars using SVCS. However, image formation alone is not sufficient; we must also reliably extract scintillation parameters from the visibility data. These measurements are highly sensitive to the structure of the RFI, and the noise in dynamic spectra is typically $\sqrt{N}$ higher than in the corresponding radio images, where N is the product of channels and time samples. This requires more aggressive flagging. To mitigate RFI, we exclude all central square baselines, which are particularly prone to contamination. For the estimation of the scintillation parameters, we process each target scan individually. Although this approach reduces the overall SNR, it enables more accurate and confident measurement of scintillation properties. 

We have developed SPICE (Scintillation Pipeline for Interferometric Candidate Extraction), a fully automated CASA‑based pipeline designed to identify pulsar candidates in GMRT and uGMRT interferometric data through their scintillation signatures. By integrating robust flagging, iterative calibration, adaptive imaging, and automated classification, SPICE provides a reproducible framework for extracting scintillation parameters from large archival datasets. Our tests on legacy GMRT and uGMRT observations demonstrate its ability to recover known pulsars and quantify their scintillation properties, while also highlighting challenges posed by RFI, reference antenna dependence, and intrinsic variability of scintillation parameters. 

The performance of SPICE is ultimately constrained by telescope sensitivity, bandwidth, and time–frequency resolution, which limit the measurement of true scintillation bandwidths and timescales. Nevertheless, the pipeline complements time‑domain searches by enabling image‑based identification of compact variable sources. Its scalability across data volumes and adaptability to different observing conditions makes it a valuable tool for exploiting the extensive GMRT archive and for future surveys with wideband interferometers.

Incorporating higher‑resolution baseband data, improved RFI mitigation, and machine‑learning‑based classification could further improve SPICE’s efficiency and reliability. By enabling the detection of challenging pulsar populations such as pulsars in compact, highly accelerated binary orbits, pulsars near the Galactic Center, or aligned rotators, SVCS can play a key role in expanding the known pulsar population and probing the turbulent interstellar medium.
\bibliography{sample631}{}
\bibliographystyle{aasjournal}

\end{document}